# Acoustic Prism for Continuous Beam Steering Based on Piezoelectric Metamaterial

J. Xu and J. Tang
Department of Mechanical Engineering
University of Connecticut
Storrs, CT 06269, USA
Phone: (860) 486-5911, Email: jtang@engr.uconn.edu and jia.xu@uconn.edu

## ABSTRACT

This paper investigates an acoustic prism for continuous acoustic beam steering by a simple frequency sweep. This idea takes advantages of acoustic wave velocity shifting in metamaterials in the vicinity of local resonance. We apply this concept into the piezoelectric metamaterial consisting of host medium and piezoelectric LC shunt. Theoretical modeling and FEM simulations are carried out. It is shown that the phase velocity of acoustic wave changes dramatically in the vicinity of local resonance. The directions of acoustic wave can be adjusted continuously between 2 to 16 degrees by a simple sweep of the excitation frequency. Such an electro-mechanical coupling system has a feature of adjusting local resonance without altering the mechanical part of the system.

## 1. BACKGROUND

Structural health monitoring (SHM) aims at utilizing autonomous damage detection strategies to monitor a structure in real time. A SHM system consists of sensors and actuators with data acquisition, computation and signal interpretation modules. Many types of actuators and sensors have been adopted including, for example, the piezoelectric transducers, fiber composites, magnetostrictive materials and fiber optics. Among these, piezoelectric transducers have received significant attention due to its compactness, wide bandwidth, and good linearity within functional range [1-3]. The piezoelectric transducers are commonly used in the impedance based and the guided wave based SHM systems. Towards the impedance based method, electrical impedance of the piezoelectric transducer bonded to structure is measured. Due to the two-way electromechanical coupling effect, the damage induced mechanical impedance changes can be reflected in the electrical impedance shifting [4, 5]. The impedance based method has advantages of simple configuration, low power consumption, and independent on an analytical model for implementation [4-7]. Alternatively, the guided wave propagation approach shows better feasibility in far field damage detection [8-10]. This is because the guided Lamb waves have the advantage of wave propagation over long distances with little loss of amplitude. Consequently, it is not required to place the actuators or sensors in the vicinity of damages. By extending the guided wave method into 2- or 3-dimensional space, the phased array techniques arise and have been subject to wide exploration. In the phased array method, the surface bonded actuators focus the wave energy on localized directions or areas by controlled excitation time delay of each array element [11-15]. The beam forming direction reorientation can be achieved by altering the excitation time sequences. Nevertheless, this method needs complicate and precise control coordination. Due to the demand of low cost acoustic beam steering, a frequency-based beam steering method has been proposed in a recent study by utilizing periodic array of piezoelectric actuators [16]. Piezoelectric actuators are arranged in rectangular array with distance of one wavelength between each other. The proposed array can generates strong, frequency dependent directional beaming, and therefore allows beam steering through a simple sweep of the excitation frequency. On the other hand, only limited wave propagation directions can be obtained due to constrain of the geometry configuration. Real-time geometrical adjustment of the frequency-based beam steering device might be difficult.

## 2. RESEARCH OVERVIEW

It has been known in optics that light can be dispersed through a prism. More specifically, light beams with difference frequencies have different dispersion angles when travelling through a prism. Inspired by the working principle of optical prism, the hypothesis of this research is that an acoustic prism may have similar feature, that is, steering acoustic wave continuously by shifting the frequency of an acoustic wave.  There are mainly two challenges to build such an acoustic prism.  One challenge here is that we need significant velocity change of acoustic wave.  This is because that the refraction angle of a wave is dependent of the difference of phase velocities of two mediums.  The other challenge is that we need phase velocity change within a small frequency range for better controllability in application.

Acoustic metamaterial shows good feasibility to build such a prim.  The metamaterial, defined as artificial structures that exhibit physical properties not available in natural material, has extraordinary capability in low-frequency sound/vibration attenuation, negative refraction, and acoustic/elastic lenses [17-24].  Note that phase velocity of an acoustic wave in metamaterial undergoes significantly shifting in the vicinity of local resonance [25, 26].  Here we propose an acoustic prism for beam steering.  By utilizing the phase velocity shifting, the direction of acoustic wave can be adjusted continuously by a simple sweep of the excitation frequency.  We apply this concept into the metamaterial with piezoelectric shunt circuit.  Our rationale here is that velocity of acoustic wave would shift dramatically in the vicinity of the LC resonance which further yields the beam steering effect.  It is worth mentioning that such an electro-mechanical coupling system has a unique feature of adjusting local resonance without altering the mechanical part of the system.  The full integration of adaptive materials, electronics, computing resources, and power systems with passive metamaterials can form a hybrid active metamaterial system whose material properties can be digitally and remotely controlled.

The rest of this paper is organized as follows. Section 3 describes the concept of the acoustic prism and presents the modeling and analysis of directional guided wave excitation in thin plates.   Section 4 presents simulated validations of the concept, and Section 5 summarizes the main results of the study and provides recommendation for future investigations.

## 3. PIEZOELECTRIC METAMATERIAL BASED ACOUSTIC PRISM

In this section, we formulate analytical investigation on the piezoelectric metamaterial based prism.  We first describe the basic idea, followed by a general mathematical model of the integrated system with illustration.

### 3.1 Concept of metamaterial based prism

The acoustic prism takes advantages of wave refraction.   Refraction is the bending of a wave when it enters a medium where its speed is different.   The refraction phenomenon exists widely in optics, electrics and acoustic.   For example, the refraction of light when it passes from a fast medium to a slow medium bends the light ray toward the normal to the boundary between the two media (Figure 1).

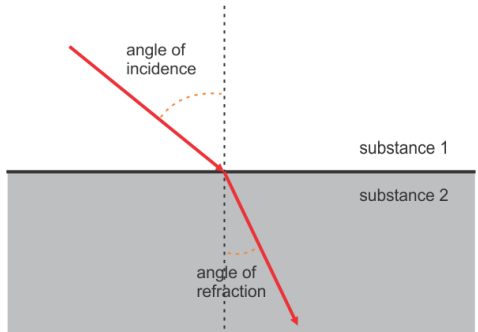

Figure 1. Illustration of refraction.

The refractions of wave normally have the following relation:

$$\frac{1}{c_{p1}(\omega)}\sin\theta_1 = \frac{1}{c_{p2}(\omega)}\sin\theta_2 \qquad (1)$$

where $c_{p1}(\omega)$ and $c_{p2}(\omega)$ are the phase velocities of the wave in the plate and the prism, respectively; $\theta_1$ and $\theta_2$ are the angle of incidence and refraction, respectively.  The relation of the in-angle and the refraction angle can be rewrite as:

$$\theta_2 = \arcsin\left(\frac{c_{p2}(\omega)}{c_{p1}(\omega)}\sin\theta_1\right) \qquad (2)$$

The phase velocity $c_{p1}(\omega)$ in the plate and the incidence angle $\theta_1$ are constant when the parameters of the system are chosen. On the other hand, the phase velocity in the prism $c_{p2}(\omega)$ can be modified subjected to the frequency-dependent material characteristic. For example, the phase velocities of the light beam shifts for the light with different colors. This phenomenon yields an interesting optical element called prism, which can disperse the light beams in different directions (Figure 2a). Note that the phase velocity of acoustic wave can be easily modified in a metamaterial due to local resonance effect. Here we propose a design of acoustic prism which can disperse the acoustic wave. The dispersion effect ultimately yields acoustic beam steering continuously.

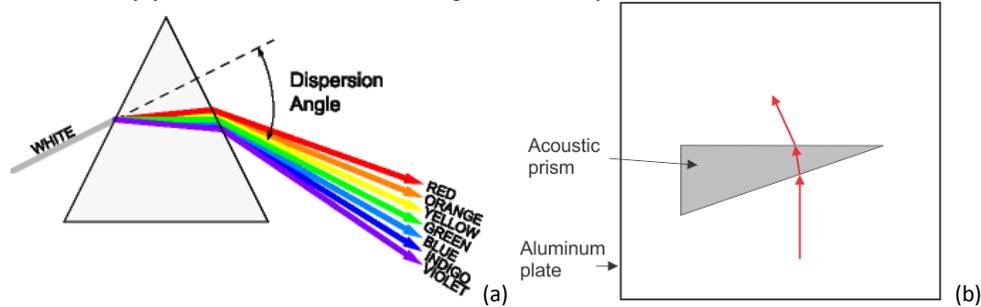

Figure 2. (a) Optical prism; (b) prototype of the acoustic prism.

The configuration of the metamaterial based prism is shown in Figure 2b. The prototype consists of unit cells arranged in triangle array to form an acoustic prism. The prism is bonded to a host plate made of aluminum sheet. The unit cell consists of the host structure and a piezoelectric transducer. An inductive shunt circuit is connected to the piezoelectric transducer to create the local resonance. A stand-alone piezoelectric transducer is placed near the acoustic prism as an acoustic wave source. Due to the local resonance, the triangle prism area may have significant acoustic velocity change in the vicinity of the local resonance. In such a case, the acoustic wave travelling through the prism may have difference refraction angle. Therefore, the direction of the acoustic wave can be easily changed by frequency sweep. Moreover, such an electro-mechanical coupling system has a unique feature of on-line tunability, that is, adjusting local resonance without altering the mechanical part of the system.

### 3.2 Mathematical model

In this section, an electro-mechanical modeling is presented to illustrate the responses of the piezoelectric metamaterial. For simplicity and without loss of generality, here we adopt the two-dimensional mass-spring lattice, as shown in Figure 3a. This model consists of periodic microstructures with host medium of mass $M$. We refer the $x$ and $y$ directions as the principal directions. The stiffness of the host medium is given by the extensional spring $K$ and the shear spring $G$. We assume that a piezoelectric transducer is bonded to the host medium. An inductor is connected to the transducer (Figure 3b). Therefore, this lattice model can be used to represent a composite material with distributed piezoelectric shunts.

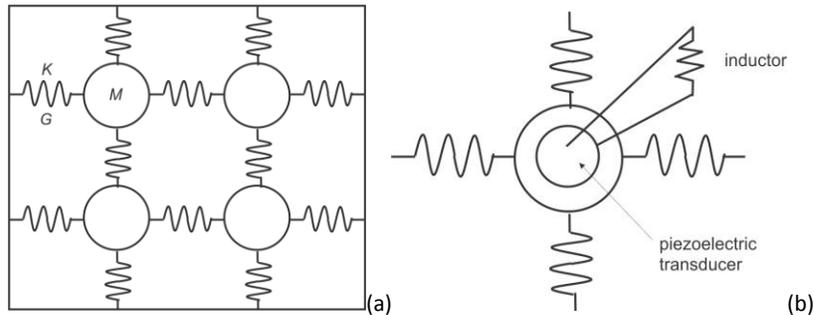

Figure 3. (a) 2D lattice system; (b) representative unit cell.

Consider a lattice point at location ($xi$, $yi$) with reference to the principal coordinates $x$ and $y$. The displacement in $x$ direction of the $i$th host medium $M$, which is assumed to be rigid, is denoted as $u_{xi}$; the displacement in $y$ direction is denoted as $u_{yi}$. The equations of motion for this unit cell/microstructure are

$$M\ddot{u}_{xi} + 2(K+G)u_{xi} + k_1 Q = K(u_{xi+1} + u_{xi-1}) + G(u_{yi+1} + u_{yi-1})$$
$$M\ddot{u}_{yi} + 2(K+G)u_{yi} + k_1 Q = K(u_{yi+1} + u_{yi-1}) + G(u_{xi+1} + u_{xi-1}) \quad (3)$$
$$L\ddot{Q} + k_2 Q + k_1(u_{xi} + u_{yi}) = 0$$

where $Q$ is the electrical charge on the surface of the piezoelectric transducer, $k_2$ is the inverse of the capacitance of the piezoelectric transducer, $R$ represents the resistance value of the shunt circuit and $L$ indicates the inductor in the shunt circuit. Let $l$ represents the length of the unit cell. The solution for a plane harmonic wave in an infinite lattice system is given

$$u_{xi} = u_{x0} e^{i(-\omega t + k_x x + k_y y)}$$
$$u_{yi} = u_{y0} e^{i(-\omega t + k_x x + k_y y)} \quad (4)$$
$$Q = Q_0 e^{i(-\omega t + \beta)}$$

We substitute Equation (4) into (3). For a nontrivial solution for wave amplitudes $u_{x0}$ and $u_{y0}$, terms associated with time must vanish. This leads to the dispersion equation, which can be solved for the wave frequency for given values of dimensionless wave numbers $k_x l$ and $k_y l$. The dispersion equation for a unit cell with LC shunt circuit is given as

$$\left(-\omega^2 L + k_2\right) - k_1^2 \frac{\left(-\omega^2 M + 2K(1-\cos k_y l) + 2G\right) + 2G\cos k_y l + \left(-\omega^2 M + 2K(1-\cos k_x l) + 2G\right) + 2G\cos k_x l}{\left(\left(-\omega^2 M + 2K(1-\cos k_y l) + 2G\right)\left(-\omega^2 M + 2K(1-\cos k_x l) + 2G\right) - 4G^2 \cos k_x l \cos k_y l\right)} = 0 \quad (5)$$

All propagating modes for lattice models can be captured by restricting the dimensionless wavenumber to the first Brillouin zone due to periodicity. Thus, we plot all the dispersion curves based on wavenumbers within this zone. In the following theoretical analysis, we choose the parameters as shown in Table 1.

TABLE I. Parameters.

| | |
|---|---|
| Mass (kg/m) | 0.08 |
| Spring constant (GN/m) | 21 |
| Shear spring (GN/m) | 8 |
| Lattice space (m) | 0.005 |
| Capacitance of transducer (nF) | 2.522 |
| Inductor (H) | 0.1 |
| Resonance of shunt circuit (kHz) | 10 |

Dispersion curves for the unit cell with and without LC shunt circuit are shown in Figures 4a and 4b, respectively, in the first Brillouin zone. Without of generality, here we select the resonant frequency of the LC shunt circuit to be 10 kHz and the system level electro-mechanical coupling coefficient is chosen to be 0.01. The system level electro-mechanical coupling coefficient is defined

$$k_e^2 = \frac{k_1^2}{K k_2} \quad (6)$$

The system level electro-mechanical coupling coefficient $k_e^2$ indicates the conversion efficiency between electrical and acoustic energy in piezoelectric material. The electro-mechanical coupling coefficient is directly related to the transducer material property, predominantly the piezoelectric coupling constant at the material-level. For example, one of the most commonly used piezoelectric transducers, PZT5H, has limited piezoelectric coupling constant of 0.44 in the 31 direction and 0.75 in the 33 direction, respectively. The device level electro-mechanical coupling coefficient is related to not only the material property of the transducer, but also the specific design features of the harvester. It can be modified by choosing different piezoelectric material or structural level optimization [27-30].

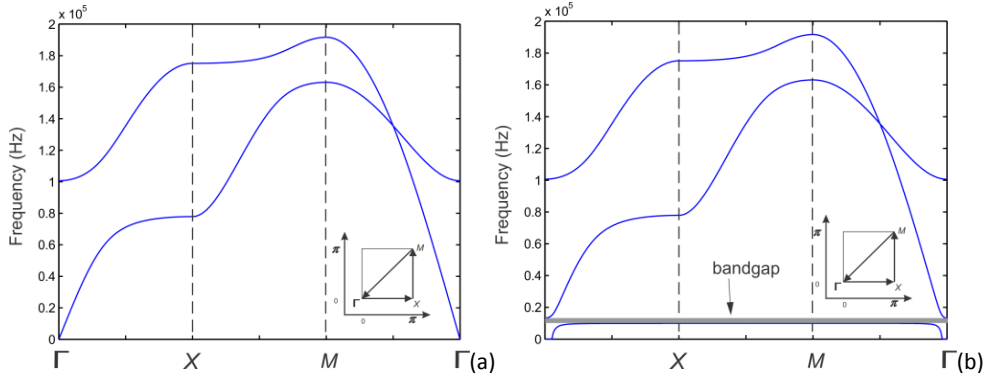

Figure 4. Dispersion curve (a) without shunt circuit; (b) with LC shunt circuit.

First, we analyze the influence of LC shunt towards the dispersion curve of the unit cell, as shown in Figures 4a and 4b, i.e., the dispersion curves of the unit cell without and with LC shunts. It can be observed that the LC shunt circuit induces a bandgap around the LC resonant frequency. As a result, the phase velocity of acoustic wave changes in the vicinity of this bandgap. The phase velocity of acoustic wave can be obtained through the dispersion curve, where

$$C_p(\omega) = \frac{\omega}{k_x} \tag{6}$$

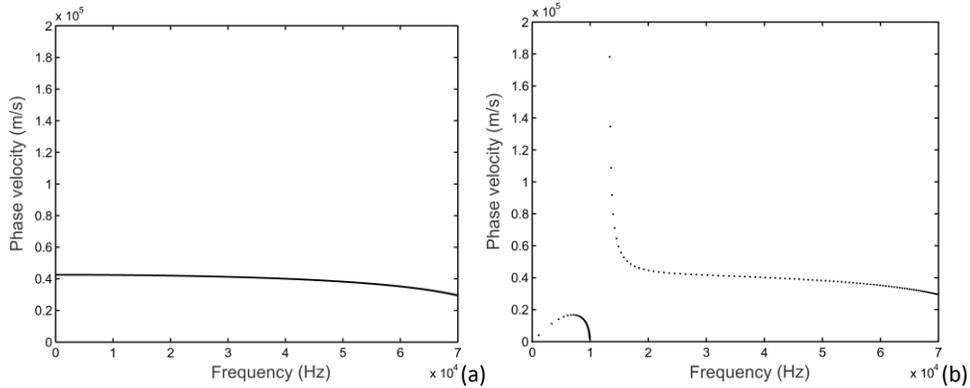

Figure 5. Phase velocity (a) without shunt circuit; (b) with LC shunt circuit.

We may also obtain respectively the phase velocity of acoustic wave within unit cell without and with LC shunt circuit, as shown in Figures 5a and 5b. It can be observed that while the phase velocity with LC shunt circuit generally follows the tendency of the one without shunts, it shifts significantly in the vicinity of the LC resonance. It indicates that the acoustic wave, when passes from the plate area into the metamaterial area, may dramatically change its direction due to the shifting of wave speed. This phenomenon further yields the frequency dependent beam steering. Moreover, we obtain continuous phase velocity shifting during the abovementioned frequency range. It indicates that the angle of the acoustic wave can be tuned continuously. This capability has an advantage over the periodic array of piezoelectric actuators [16] due to its continuous adjustability and compact size. It is also worth mentioning that, the local resonance due to LC shunt circuit is dependent upon the parameters of the capacitance and the inductor. One can easily change the frequency of the LC resonance by simply modifying the value of the inductor, i.e., without altering the mechanical part of the prism.

In the next step, we analyze the system characteristics subjected to different electro-mechanical coupling coefficients. Figures 6a to 6d show the dispersion curves of the unit cell with different system level electro-mechanical coupling coefficients. It can be observed that LC shunts induces a bandgap for all of the four cases. Moreover, increasing the coupling coefficient can effectively increase the width bandgaps. For example, the unit cell with $k_e^2 = 0.005$ has bandgap width of 13.14 kHz. On the other hand, when the electro-mechanical coupling of the unit cell is increased to $k_e^2 = 0.1$, the bandgap width is increased to 70 kHz. This is because that increasing electro-mechanical coupling coefficient may increase the portion of acoustic energy that converted to electrical energy. Subsequently the LC resonance of the shunt circuit may generate a larger reacting force that applied to the piezoelectric transducers that prevent wave propagation within the unit cell. Therefore the bandgaps can be expanded through increasing the system level electro-mechanical coupling coefficient.

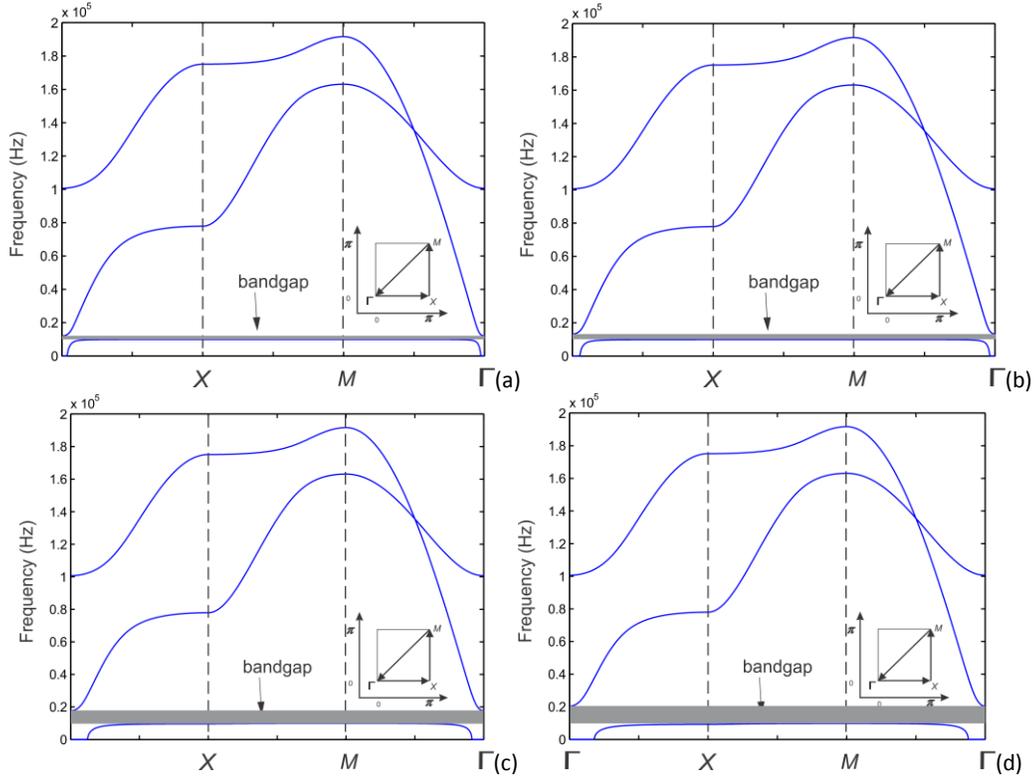

Figure 6. Dispersion curve (a) $k_e^2 = 0.005$; (b) $k_e^2 = 0.01$; (c) $k_e^2 = 0.05$; (d) $k_e^2 = 0.1$.

Figures 7a to 7d show the phase velocity of acoustic wave within the unit cell with different electro-mechanical coupling coefficient. It can be observed that, the phase velocity has significant changes in the vicinity of local resonance. The bandgap of the phase velocity expands with the increased electro-mechanical coupling coefficient. On the other hand, the electro-mechanical coupling coefficient may reduce the phase velocity of acoustic wave in low frequency range. It indicates that we may increase the refraction angle by increase the electro-mechanical coupling coefficient. Meanwhile, the phase velocity changes little due to the coupling coefficient in high frequency range.

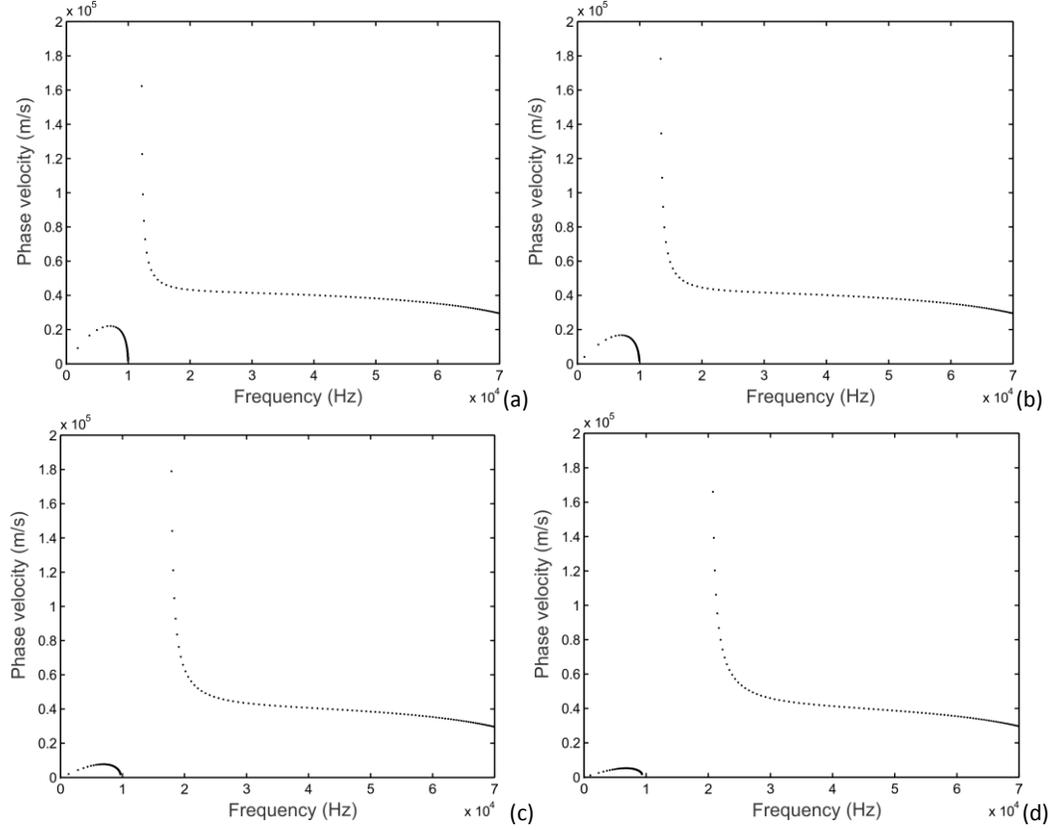

Figure 7. Phase velocity (a) $k_e^2 = 0.005$ ; (b) $k_e^2 = 0.01$ ; (c) $k_e^2 = 0.05$ ; (d) $k_e^2 = 0.1$ .

## 4. FEM SIMULATION AND DISCUSSION

Finite element method (FEM) simulations are carried out to verify the proposed concepts. The configuration of the unit cell is shown in Figure 8a. The unit cell consists of aluminum substrate (size 5×5×1 mm³) and a piezoelectric transducer (size 4×4×0.8 mm³) bonded on its surface. An inductor is connected to the top and bottom surfaces of the piezoelectric transducer as the shunt circuit. The parameters for the materials are: $\rho_p = 7500$ kg/m³ , $\rho_b = 2700$ kg/m³ , $E_b = 62$ GPa, and $E_p = 106$ GPa. The material constants of the piezoelectric transducer, PZT-5H, are $d_{31} = -320$ pC/N  $h_{31} = 5.9 \times 10^8$ N/C and  $\beta_{33} = 2.92 \times 10^8$ Vm/C.

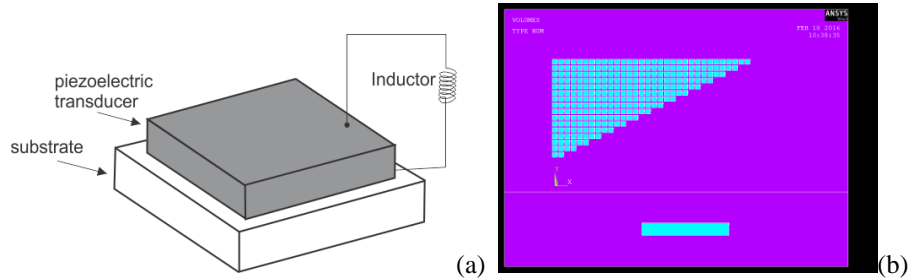

Figure 8. (a) Configuration of the unit cell; (b) Configuration of the metamaterial based prism.

The unit cells are arranged into a triangle-shaped array to form the acoustic prism, as shown in Figure 8b. The lengths of the two sides of the triangle prism are 160 and 80 mm, respectively. A stand-alone piezoelectric transducer is placed in the vicinity of the prism as the wave source. The substrate plate is made of aluminum, and the thickness of the plate is 10 mm. ANSYS 14.5 is used for the FEM simulation. In the simulation, the capacitance of the piezoelectric transducer is used as 2.522 nF. Without of generality, we choose the resonant frequency of the LC shunt circuit to be 10k Hz. Therefore, the inductor in the circuit is chosen to be 0.1392 H. Simulations are performed under excitation of different frequencies around the local resonance of the shunt circuit.

The electro-mechanical coupling coefficient $k_e^2$ is hinged upon the material parameters and system configuration, e.g., the thickness ratio and size of the piezoelectric transducer. On the other hand, active circuit can also increase the system level electro-mechanical coupling coefficient. For example, the negative capacitance circuit (Figure 9) shows extraordinary ability in the improvement of the coupling coefficient [31]. The bandwidth of the tuning range is expanded by increasing the electro-mechanical coupling coefficient. This is because that increased electro-mechanical coefficient of the piezoelectric material increases the system level electro-mechanical coupling coefficient, which yields the reduction of the anti-resonance frequency. An improved coupling coefficient may induce expanded tuning frequency range. The resulting benefit is that the direction of the acoustic beam can be turned with better resolution.

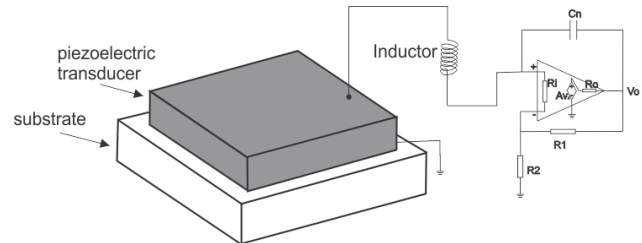

Figure 9. Unit cell with inductor & negative capacitance shunt circuit.

In the next step, FEM simulations are carried out that the negative capacitance circuit has compensated 90% of the stiffness of the piezoelectric transducer. In other words, the electro-mechanical coupling coefficient of the piezoelectric transducer is increased by 10 times for enhanced beam steering effect. Figure 10 shows representative examples of the wave steering effect

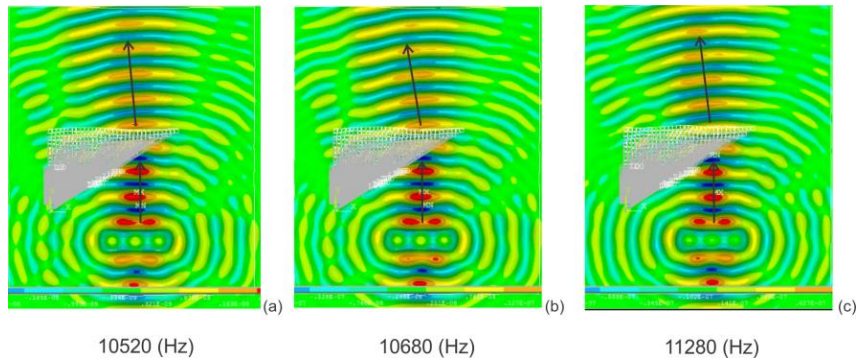

Figure 10. Illustration of beam steering effect.

It can be observed from Figure 10 that when the frequency of the wave increased from 10520 Hz to 10680 Hz, the direction of the wave is bended to the left. Moreover, when the frequency is further increased, the angle of refraction is decreased. This is because that in the vicinity of the local resonance, as aforementioned, we have significant changes of the phase velocity. The phase velocity shifting ultimately yields the change of the wave travelling direction. The relation of the steering angle and frequency of the acoustic wave is shown in Figure 11 which shows the steering angle versus the frequency of the wave. The results are obtained by FEM simulation in steady-state condition. It can be obtained that when the frequency of the wave increases from 10.5 kHz to 11 kHz, the steering angle of the wave changes from 6 to 16 degree. This is because the prism with LC shunt circuit has decrease the phase velocity of the wave dependent on the frequency. Therefore, frequency dependent wave steering effect is achieved. Moreover, due to the local resonance, the LC shunt circuit reduces the phase velocity of the wave to negative and create a bandgap zone between 11.79 to 12.3 kHz. In this zone the wave from the source has been reflected back. In the range between 12.3 and 13 kHz, the steering angle decrease from 12 degree to 1.8 degree.  The tendency of the steering angle matches the theoretical prediction in general.

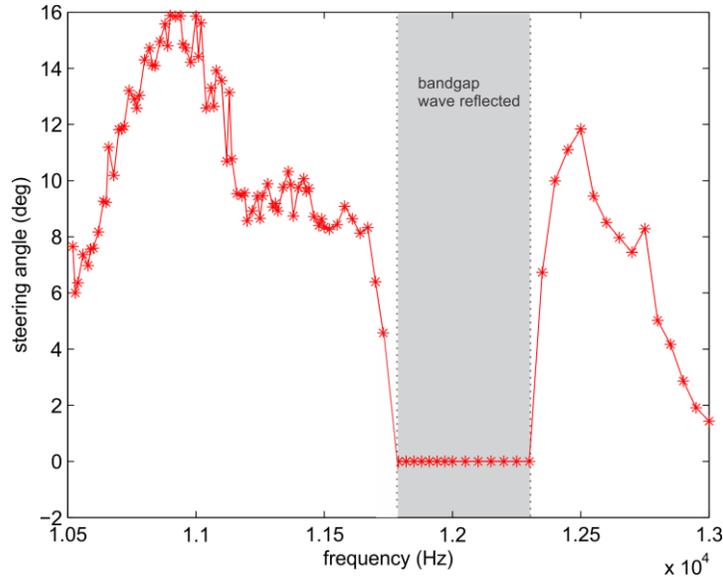

Figure 11. Steering angle versus frequency of the wave.

## 5. CONCLUDING REMARKS

Our research finds that the proposed metamaterial based prism offers the capability of steering the acoustic wave due to the local resonance from the LC shunt circuit.   Here we have illustrated the benefits of the capability of acoustic beam steering by choosing the resonance of the LC shunt circuit to be 10 kHz.   The angle of the wave can be tuned between 2 to 16 degrees in our simulation.   It is worth noticing that the resonant frequency of the shunt circuit can be modified by changing the value of the inductor without mechanical tailoring.   For example, we may utilize a simulated inductor with on-line controllability.   The concept proposed here simplifies the configuration and control strategy of wave steering.   It can be applied in structural health monitoring and ultra-sonic wave generators.

## 6. ACKNOWLEDGMENT

This research is supported in part by NSF under grant CPS – 1544707.